\documentclass[twocolumn,showpacs,preprintnumbers,amsmath,amssymb]{revtex4}

\usepackage{graphicx}

\begin{document}
\title{Ultracold mixtures of atomic $^{6}$Li and $^{133}$Cs with tunable interactions}
\author{Shih-Kuang Tung}
\author{Colin Parker}
\author{Jacob Johansen}
\author{Cheng Chin}
\affiliation{The James Franck Institute and Department of Physics, The University of Chicago, Chicago, IL 60637, USA}
\date{January 5, 2013}
\author{Yujun Wang}
\author{Paul S. Julienne}
\affiliation{Joint Quantum Institute, NIST and the University of Maryland, 100 Bureau Drive Stop 8423, Gaithersburg, Maryland 20899-8423, USA}

\begin{abstract}
We report the experimental and theoretical study of two-body interactions in a $^{6}$Li-$^{133}$Cs Fermi-Bose mixture.  Using a translatable dipole trap setup, we have successfully trapped the two species in the same trap with temperatures of a few microkelvins. By monitoring atom number loss and interspecies thermalization, we identify five $s$-wave interspecies Feshbach resonances in the lowest two scattering channels. We construct a coupled channels model using molecular potentials to fit and characterize these resonances.  Two of the resonances are as wide as 6~mT (60~G) and thus should be suitable for creating Feshbach molecules and searching for universal few-body scaling.
\end{abstract}


\maketitle

In ultracold atom experiments, the interaction between two atoms can be controlled by tuning an external magnetic field near a Feshbach resonance \cite{Chin1}. The ability to control precisely the interaction between neutral atoms has opened up many possibilities for ultracold atom research. From a few-body physics perspective, it has allowed the creation of Feshbach molecules \cite{Julienne1} and Efimov trimers \cite{Efimov1,Grimm1}. From a many-body physics perspective, it has allowed exploration of BEC-BCS crossover \cite{Jin5,Grimm2,Ketterle3} and molecular Bose-Einstein condensates \cite{Grimm3,Jin4} as well as bringing a dilute gaseous system from a weakly interacting regime to a strongly interacting regime\cite{Wieman1,Thomas1}.

Using a heteronuclear mixture opens up even more possibilities. By mixing bosonic and fermionic atomic species, one can create molecules with different statistics. A mixture of bosonic and fermionic quantum gases \cite{Ketterle2,Inguscio1} is also interesting as it can provide a new test ground for many-body physics. In particular, marrying  the mixture with optical lattices, another powerful and tunable tool, the mixture provides new opportunities to explore exotic quantum phases, such as supersolids \cite{Blatter1}, insulators with fermionic domains \cite{Lewenstein1}, and boson mediated superfluids \cite{Efremov1}. Furthermore, an ultracold heteronuclear mixture can lead to ground-state molecules with a large electric dipole moment and high phase space density \cite{JY1,JY2} to study quantum gases with long-range dipolar interactions.

A key ingredient of the above research prospects is the control of the interspecies interaction. In this Rapid Communication, we report the production of an ultracold mixture of fermionic $^{6}$Li and bosonic $^{133}$Cs atoms and the observation of five interspecies Feshbach resonances. Our choice of the Li-Cs mixture provides unique opportunities to explore new quantum physics. First, heavy Cs atoms can be cooled down to $\approx$10 nK \cite{Chin2,Chin3}, while a Fermi gas with lighter lithium atoms can have $T_F \approx 1 \textrm{ }\mu\textrm{K}$. Effective sympathetic cooling with Cs atoms would prepare Li atoms at a temperature of $T$/$T_{F}$=0.01, crucial for observations of exotic quantum phases \cite{Bloch1,Lukin1,Esslinger1}. Second, the distinct optical  excitations of $^{6}$Li and $^{133}$Cs allow for independent manipulation of the two species in a bichromatic optical lattice for quantum information processing \cite{Chin4}. Third, the extreme mass ratio of the two species leads to a small universal scaling factor of $\approx$5 for the LiCs$_2$ Efimov trimer states \cite{Esry1,Petrov1}, compared to 22.7 for a homonuclear system \cite{Efimov1,Grimm1}. A smaller scaling constant facilitates the test of the scaling law in Efimov physics. Finally, ground-state LiCs molecules possess the largest dipole moment of 1.8$\times$10$^{-29}$ C~m (5.5 D) \cite{Aymar,Weidemuller2} that one can obtain by combining two stable alkali-metal atoms. 

There are, however, challenges in the laboratory preparation of the Li-Cs mixture, including a disparity in the sample temperatures after optical cooling ($\approx$300 $\mu$K for Li and $\approx$2 $\mu$K for Cs), and the large inelastic collisions between optically excited Cs atoms and ground-state lithium atoms \cite{Weidemuller1}. To avoid the undesired heating and loss when cold Cs atoms are loaded into a very deep trap ($\approx$1~mK) and the light-assisted inelastic collision loss when Li atoms overlap with a Cs magneto-optical trap (MOT), we design a translatable dipole trap. This trap allows us to evaporate Li atoms first, lower the dipole trap power, and then shift Li atoms to a different location prior to the Cs MOT loading. 

\begin{figure}[t]
\includegraphics[width=3.2 in]{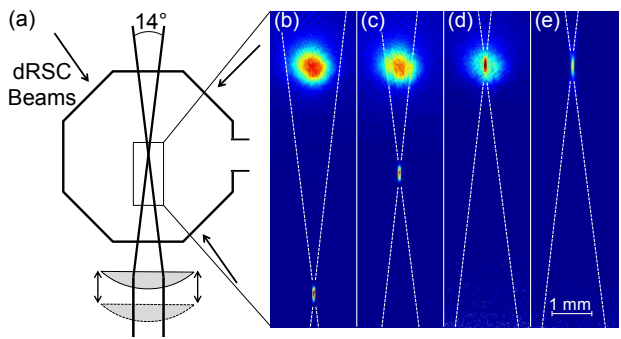}
\caption{(Color) Optical setup and procedure for combining $^{6}$Li and $^{133}$Cs clouds: (a) Schematic for the translatable crossed dipole trap. The position where the two dipole beams cross can be shifted by moving the lens mounted on a motorized translation stage. (b)--(e) Top view images show the merging process for the two atomic clouds. The $^{6}$Li atoms are initially trapped in an offset crossed dipole trap, while the $^{133}$Cs atoms are loaded and cooled in the optical lattice of the degenerate Raman-sideband cooling (dRSC). After the two clouds are merged, both species are confined in the crossed dipole trap. Dotted lines show the dipole beam paths.} \label{fig1}
\end{figure}

Our experiment starts by loading the Li MOT for 6 s to collect 10$^{8}$ Li atoms from the Zeeman slowed atomic beam. After an 8-ms compressed MOT phase, we pump Li atoms into the $F$~=~1/2 manifold, during which we collect 10$^{6}$ atoms into a translatable dipole trap. The dipole trap is implemented by focusing two laser beams through a lens mounted on a motorized translation stage; see Fig.~\ref{fig1}. The two laser beams cross at an angle of 14$^{\circ}$ with a beam diameter of 110 $\mu$m at the focus. A diode-pumped ytterbium fiber laser at a wavelength of $\approx$1070 nm provides a maximum laser power of 70 W in each beam. The dipole trap has a maximum trap depth of 1.3 mK for Li atoms and gives radial and axial trapping frequencies ($\omega_r$,~$\omega_a$) = 2$\pi\times$ (7, 0.84) kHz. With 3 s of Li evaporation at a magnetic field of 830~G, the temperature of the Li atoms is lowered to $\approx$3~$\mu$K and 2$\times$10$^{5}$ Li atoms remain in the trap. At this stage, the power in each dipole beam has decreased to 0.64~$\%$ of its full power, which reduces the trap depth by the same factor and reduces the trapping frequencies by a factor of 12.5. We then increase the trap depth slightly, and move the crossed dipole trap 12 mm away from the MOT position before loading Cs atoms. We are able to move the Li cloud over 12 mm in 300 ms without any observable Li number loss and heating. After loading Cs atoms for 5 s in a MOT, we obtain 5$\times$10$^{7}$ Cs atoms.  We then perform a magnetic compression, molasses cooling and degenerate Raman-sideband cooling \cite{Chu1}. After all optical cooling, we move the crossed dipole trap back to the original position and release the Cs atoms to combine the two species ( see Fig.~\ref{fig1}). Subsequently, we have 10$^{5}$ Li atoms and 2$\times$10$^{5}$ Cs atoms with temperatures of 3 and 10 $\mu$K, respectively.  We do not observe clear thermalization between the two species at low magnetic field.

In the dipole trap, we can independently prepare Li and Cs atoms in different spin states. Following the convention in Ref. \cite{Chin1}, we label the hyperfine energy levels alphabetically. For example, a $\left| \textrm{Li:} b \right>$ and $\left| \textrm{Cs:} a \right>$ mixture means Li atoms are in the second-lowest-energy state $\left| b \right>=\left| F = 1/2, m_F=1/2 \right>$ and the Cs atoms are in the lowest-energy state $\left| a \right>=\left|F=3,m_F=3\right>$.

\begin{figure}[t]
\includegraphics[width=3.3 in]{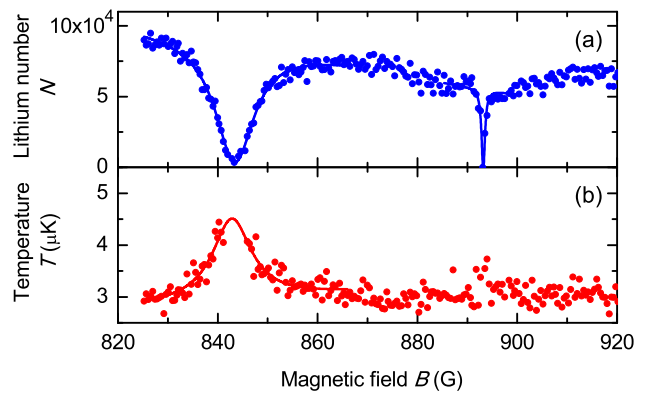}
\caption{(Color online) The Feshbach loss spectrum for the open channel of $\left| \textrm{Li:} a \right>$ and $\left| \textrm{Cs:} a \right>$. (a) Number and (b) temperature of Li atoms when combined with Cs atoms for 50 ms at various fields. The Li atom number decreases dramatically while the temperature increases when the field is tuned near resonance. Two resonances are found at 843.4 and 892.9 G. The solid lines are Lorentzian fits to the data. Each data point is averaged over three experimental realizations.} \label{aResonances}
\end{figure}

We prepare atoms in the desired spin states and perform Feshbach spectroscopy in the dipole trap. We first perform spectroscopy on a $\left| \textrm{Li:} a \right>$-$\left| \textrm{Cs:} a \right>$ mixture.  To purify the spin of the Li atoms, we apply a resonant laser pulse to push the $\left| \textrm{Li:} b \right>$ atoms out of the trap. The spin cleaning process for Li is implemented at 430~G \cite{Gauss}, where the optical transitions from $\left| \textrm{Li:} a \right>$ and $\left| \textrm{Li:} b \right>$ to the 2$P_{3/2}$ states can be resolved. Cs atoms have a more complex spin structure. To spin polarize Cs atoms into $\left| \textrm{Cs:} a \right>$, we apply a microwave pulse with a frequency chirp that transfers atoms in unwanted sublevels of the $F$ = 3 manifold to sublevels in the $F$ = 4 manifold, then remove them by a resonant laser pulse. 

We choose to measure the trap loss and temperature of Li atoms as the indicator of an interspecies Feshbach resonance, exploiting the fact that identical fermions do not interact at such a low temperature. Near a Li-Cs Feshbach resonance, however, Li atoms experience enhanced trap loss and heating. The heating is due to the fact that Cs atoms are hotter, and near a Feshbach resonance, the large scattering length results in faster thermalization. Since Li atoms see a shallower trap, the heating also induces fast evaporative loss. Trap loss can also come from three-body recombination between one Li and two Cs atoms, which drives the atoms out of the trap. After ramping the magnetic field in 7~ms to a desired value and holding for 50~ms, we measure the Li number by absorption imaging. For a  $\left| \textrm{Li:} a \right>$-$\left| \textrm{Cs:} a \right>$ mixture, we scan the magnetic field from 0 to 1000~G with steps of 1~G and observe two resonances located at 843.4 and 892.9~G; see Fig.~\ref{aResonances}. Fitting the loss features to a Lorentzian profile, we determine the full width at half maximum $\delta$ to be 5.6 and 1.0 G, respectively. Temperatures of Li atoms are determined by time-of-flight images. Figure ~\ref{aResonances}(b) shows that the temperature of the Li atoms rises near the resonances. The temperature increase is less obvious near the narrow resonance, since the scattering length increases within small field range. We confirmed that neither heating nor trap loss occurs when $^{133}$Cs atoms are absent. 

\begin{figure}[t]
\includegraphics[width=3.3 in]{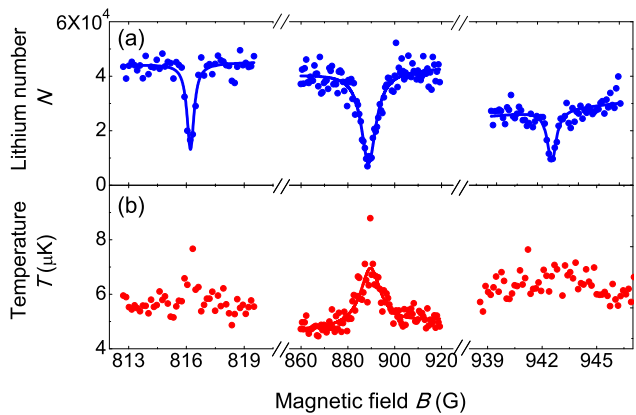}
\caption{(Color online) The Feshbach spectrum for the open channel of $\left| \textrm{Li:} b \right>$ and $\left| \textrm{Cs:} a \right>$. (a) Number and (b) temperature of Li atoms when combined with Cs atoms for 50 ms at various fields. The Li atom number decreases dramatically while the temperature increases when the field is tuned near resonance. Three resonances are found at 816.1, 889.0, and 943.4 G, respectively. The solid lines are Lorentzian fits to the data.  Each data point is averaged over three experimental realizations.} \label{bResonances}
\end{figure}

We also identify the Feshbach resonances with Li atoms in $\left| \textrm{Li:} b \right>$ and Cs atoms in $\left| \textrm{Cs:} a \right>$.  To prepare Li atoms in $\left| \textrm{Li:} b \right>$, we use a radio-frequency adiabatic rapid passage to transfer Li atoms from $\left| \textrm{Li:} a \right>$ to $\left| \textrm{Li:} b \right>$, then optically remove the remaining $\left| \textrm{Li:} a \right>$ atoms. The result is shown in Fig.~\ref{bResonances}. Three loss peaks are found at 816.1, 889.0, and 943.4~G, and the fitted widths $\delta$ are 0.4, 7.7, and 0.8~G, respectively. All measured Feshbach resonances are summarized in Table ~\ref{table1}. The experiment uncertainties reflect the resolution of our magnetic field calibration based on Cs microwave spectroscopy.

\begin{table}
\caption{Summary of Feshbach resonances found for $\left| \textrm{Li:} a \right>$ + $\left| \textrm{Cs:} a \right>$ and $\left| \textrm{Li:} b \right>$ + $\left| \textrm{Cs:} a \right>$. The experimental width $\delta$ is obtained by fitting the number loss data to a Lorentzian, and is not the same as the resonance width $\Delta_i=B_{0,i}-B_{x,i}$.}
\begin{center}
\begin{tabular*}{\columnwidth}{ l cccc r@{.}l c r@{.}l cccc r@{.}l c r@{.}l c c c r@{.}l}
\hline\hline
 Spin& &&&&\multicolumn{5}{c}{Experiment} &&&&&\multicolumn{10}{c}{Theory}\\
\cline{6-10} \cline{15-24}
 state &&&& &\multicolumn{2}{c}{$B_{0}$(G)} & &\multicolumn{2}{c}{$\delta$(G)} &&&&&\multicolumn{2}{c}{$B_{0}$(G)} & &\multicolumn{2}{c}{$\Delta$(G)} && $m_f$&&\multicolumn{2}{c}{$s_{res}$}\\
\hline
 $\left| \textrm{Li:} a \right>$+ &&&&& 843&4(2)&    & 5&6    &&&&& 843&1(2)&  & 62&0&  &7/2& &0&74\\
 $\left| \textrm{Cs:} a \right>$&&&& &892&9(2) &   & 1&0       &&&&& 893&0(2)&  & 2&0&  &7/2&&0&02\\
 \\
  $\left| \textrm{Li:} b \right>$+ &&&&& 816&1(2)&     & 0&4    &&&&& 816&4(2)& & 2&0& &5/2&&0&03\\
 $\left| \textrm{Cs:} a \right>$&&&&& 889&0(2)&     & 7&7    &&&&&  888&8(2)& & 59&5& &5/2 && 0&71\\
                                                 &&&&& 943&4(2)&     & 0&8    &&&&& 943&4(2)& & 2&0& &5/2& &0&02\\
\hline\hline
\end{tabular*}
\end{center}
\label{table1}
\end{table}

Using the LiCs molecular potential energy curves for the ground $X^1\Sigma_g^+$ and $a^3\Sigma_u^+$ states~\cite{Staanum}, we have constructed a standard coupled channel model ~\cite{Chin1} for calculating the bound and scattering states of Li $+$ Cs collisions for any combination of spin states of the two atoms.  Following our standard procedure \cite{Leo} of making a small harmonic variation to the inner wall of each potential in the region $R<R_e$, where $R_e$ is the equilibrium internuclear distance of the potential, we vary the potentials to modify the respective $X^1\Sigma_g^+$ and $a^3\Sigma_u^+$ scattering lengths $a_1$ and $a_3$ to fit the observed $B$ field for each of the five measured Feshbach resonances. The predicted pole position $B_0$ and width $\Delta$ of each resonance are summarized in Table~\ref{table1}.  The scattering length calculated from the coupled channel model can be fitted by $a(B) =a_\mathrm{bg} \prod_i\left[ (B-B_{x,i})/(B-B_{0,i}) \right ]$ \cite{Lange} with an accuracy of 98~$\%$ or better in the range of 750--1000~G. Here $B_{0,i}$ is the $i$-th pole, $ B_{x,i}$=$B_{0,i}-\Delta_i$ is the $i$-th zero, the background scattering length is $a_{bg}$=~--29.2$a_0$ for the $|\textrm{Li}:a\rangle$ + $|\textrm{Cs}:a\rangle$ channel, and is $a_{bg}$=~--29.3$a_0$ for the $|\textrm{Li}:b\rangle$ + $|\textrm{Cs}:a\rangle$ channel. The scattering lengths for our best fit model are $a_1$=~30.2(1)$a_0$ and $a_3$ =~--34.5(1)$a_0$. The error estimate represents one standard deviation in the least-squares fit to the data.  All relevant Feshbach molecular states are mixtures of singlet and triplet states, and the only good quantum number is $m_f$, the projection of the total angular momentum. 

\begin{figure}[t]
\includegraphics[width=3.2 in]{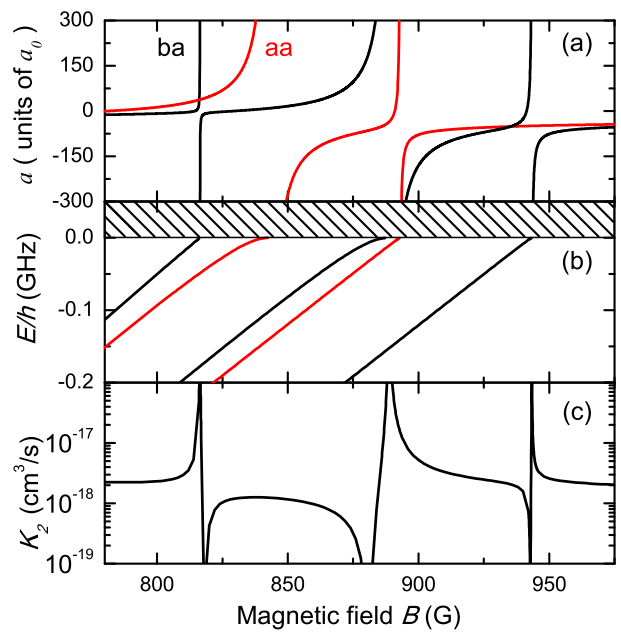}
\caption{(Color) Theoretical values for (a) scattering lengths, (b) molecular energies, and (c) the decay constant of the $|\textrm{Li}:b\rangle$ + $|\textrm{Cs}:a\rangle$ channel (black) to the $|\textrm{Li}:a\rangle$ + $|\textrm{Cs}:a\rangle$ channel (red) at a temperature of 1 $\mu$K. The shaded area in (b) indicates the scattering continuum.} \label{Theory_aB}
\end{figure}

Figure~\ref{Theory_aB} shows the scattering length variation for the two spin channels investigated and the two-body inelastic decay coefficient $K_2(B)$ for the higher-energy spin channel, which can decay to the lower one through spin-dipolar coupling to exit channel $d$ waves.  Because the $d$-wave barrier in the lower exit channel is 10.6 mK, which is much larger than the energy release in a $|\textrm{Li}:b\rangle$ to $|\textrm{Li}:a\rangle$ spin-flip transition, the loss coefficient $\approx$10$^{-18}$ cm$^3/$s away from resonance is negligible.  Thus, large changes in scattering length are possible for the $|\textrm{Li}:b\rangle +|\textrm{Cs}:a\rangle$ channel without significant decay. The channel can be used to achieve dual quantum degeneracy, as long as the detuning $|B-B_0| \gg \gamma_B$, where $\gamma_B = 3$ mG corresponds to the decay width of the closed channel component of the resonance.

From our calculation, we found the dimensionless resonance strength parameter $s_{\textrm{res}}$ \cite{Chin1} for the two broad resonances at 843 and 889 G are both slightly less than unity (see Table~\ref{table1}), indicating a significant open channel contribution. Also, there will be no other $s$-wave bound states at fields smaller than 4000~G, since the next bound states that can cross the scattering threshold are bounded by $\approx$20 GHz at zero field. 

In conclusion, we have realized an ultracold atomic mixture of $^{6}$Li and $^{133}$Cs and identify five inter-species $s$-wave resonances in two spin configurations of Li and Cs atoms. All five resonances are confirmed by theoretical calculations based on a standard coupled channel model \cite{Chin1}. The broad resonance in the $\left| \textrm{Li:} b \right>$ + $\left| \textrm{Cs:} a \right>$ channel is located in an accessible range where both a Li Fermi gas and a Cs condensate can be stable \cite{Ferlaino}. The theoretical calculations suggest that the inelastic collision rates between atoms in $\left| \textrm{Li:} b \right>$ and $\left| \textrm{Cs:} a \right>$ is low enough to  accommodate dual quantum degeneracy. Both facts suggest promising future applications of Li-Cs mixtures to study tunable Bose-Fermi gas, ultracold Feshbach molecules, and Efimov trimers. To reach dual quantum degeneracy, sympathetically cooling Li atoms by evaporating Cs atoms are under investigation. 

\textit{Note added in proof.} Recently, we became aware of a similar work by M. Weidem\"{u}ller group at Heidelberg University \cite{Wed2}. The measured s-wave resonance positions from both experiments are in excellent agreement.

We thank N. Gemelke and K.-A. B. Soderberg for their help in early development of the experiment. We acknowledge support from the NSF-MRSEC program,  NSF Award No. PHY-1206095, and AFOSR-MURI.


\begin{thebibliography}{99}
\bibitem{Gauss} Units of Gauss rather than Tesla, the accepted SI unit of magnetic field, are used in this Rapid Communication to conform to the conventional usage of this field.
\end{thebibliography}
\end{document}